\documentclass{article}

\usepackage{graphicx}

\author{Aron C. Wall\footnote{aronwall@umd.edu}
\\ \textit{Maryland Center for Fundamental Physics} \\ \textit{Department of Physics} \\ \textit{University of Maryland} \\ \textit{College Park, MD 20740-4111, USA} }
\title{A proof of the generalized second law for rapidly-evolving Rindler horizons}
\date{\today}

\begin{document}

\maketitle

\begin{abstract}
The generalized second law is proven for rapidly-evolving semiclassical Rindler horizons at each instant of time, for arbitrary interacting quantum fields minimally coupled to general relativity.  The proof requires the background spacetime to have both boost and null translation symmetry.  Possible extensions to more general horizons and matter-gravity couplings are discussed.
\end{abstract}

\section{Introduction}\label{intro}

The purpose of this article is to prove the generalized second law (GSL) in the semiclassical approximation for rapidly-changing quantum fields falling across Rindler horizons.

The GSL is the hypothesis \cite{hawking75} that the generalized entropy $S_\mathrm{gen}$ of any future horizon cannot decrease as time passes, where $S_\mathrm{gen}$ is given in general relativity by the sum of the entropy outside the horizon and a quarter of the horizon area:
\begin{equation}\label{Gen}
S_\mathrm{gen} = \frac{A}{4 \hbar G} + S_\mathrm{out}.
\end{equation}
In accordance with the arguments of Ref. \cite{10proofs}, $A$ will be interpreted as the expectation value of the area, and $S_\mathrm{out}$ will be interpreted as the von Neumann entropy:
\begin{equation}\label{von}
S_\mathrm{out} = -\mathrm{tr}(\rho\,\ln\,\rho),
\end{equation}
although because the entanglement entropy of quantum fields is divergent, some sort of renormalization scheme is necessary \cite{jacobson94}.  In the case of Rindler horizons, one must subtract from Eq. (\ref{von}) the infinite entanglement entropy of the vacuum state.  So long as one is only interested in differences in the generalized entropy, this divergence should be unimportant.  (For the same reason it is not a problem that $A$ is infinite for a Rindler horizon, because only differences in area matter.)  A fully rigorous semiclassical proof of the GSL would have to specify a renormalization procedure, but in this article I will simply assume that a satisfactory procedure exists.

The GSL is a tantalizing clue about the statistical mechanics of quantum gravity, which might illuminate the nature of the fundamental degrees of freedom of spacetime. \cite{sorkin83}\cite{JP07}.  Although there are many gedankenexperiments showing that the GSL holds in particular semiclassical situations, a general proof of the GSL in semiclassical gravity will help to clarify the situation in quantum gravity.  First of all, even if we are highly confident that the GSL will turn out to be true in our universe, knowing what physical principles are necessary to prove it will help illuminate what physical principles are required for horizon thermodynamics, and therefore perhaps the underlying principles of quantum gravity statistical mechanics.  For example, does the GSL require an unbroken Lorentz symmetry \cite{EFJW07}, or does it require the particles in nature to satisfy some entropy bound \cite{bounds}, or to satisfy some energy condition \cite{energy}?  The proof presented here will require the existence of a Lorentz-invariant and translation-invariant ground state, but imposes no other conditions on the entropy or energy.  It holds for arbitrary matter interactions, so long as the matter fields are minimally coupled to gravity.

The semiclassical GSL has already been proven for small perturbations to stationary black holes, only in the sense that the final generalized entropy at the end of the process is greater than the initial generalized entropy at the end of the process \cite{10proofs}.  For example, Frolov and Page \cite{FP93} used an S-matrix to compare the generalized entropy in the asymptotic past and future of a quasi-stationary black hole.  When the small perturbation is also slowly changing with time, one can obtain the generalized entropy in the middle of the process by linear interpolation.  But for a rapidly changing process, it is unclear from previous work whether the generalized entropy might temporarily decrease during a rapidly changing process.  Thus for rapidly changing quantum fields, it has not previously been shown whether the GSL only holds globally, as a statement about initial and final equilibrium states, or infinitesimally at every moment of time.

The result in this article shows that for Rindler horizons, the generalized entropy is nondecreasing at every instant of time, so that $d S_\mathrm{gen} / dt \ge 0$.  In an instantaneous proof of the GSL, it is no longer possible to use the first law of horizon mechanics $dE = T dS$, because this law does not hold for rapid changes to a horizon.  For example the area of the event horizon may begin to increase before any energy crosses the horizon at all.  So it is necessary to find some other relation between the area of the horizon and the energy outside of it.  Instead of the first law, I will use the Raychaudhuri and Einstein equations to show that the boost energy $K$ outside of a Rindler horizon is related to the area of the bifurcation surface:
\begin{equation}\label{AK}
A = c - 8\pi G K,
\end{equation}
where $c$ is a constant independent of the time.  The fact that the vacuum state is thermal in each Rindler wedge will then be used to relate the entropy and boost energy in each wedge to a information theoretical quantity known as the relative entropy.  This quantity satisfies a monotonicity property which will turn out to imply the GSL.

Because the proof relies on the boost symmetry of the Rindler wedge, it only works for horizon slices which are (approximately) flat planes.  Thus it does not show that the generalized entropy is increasing locally at every \emph{place} and time on the horizon, $\delta S_\mathrm{gen} / \delta t \ge 0$.

This proof is also limited to small perturbations of background spacetime; it is intended as a stepping stone towards more robust results.  For reasons given in section \ref{dis}, I expect that the proof can be extended to more general situations, including arbitrary cross-sections of arbitrary horizons, and nonminimally-coupled and/or higher-curvature theories (for which there are corrections to the Bekenstein-Hawking area law \cite{WI94}).

The plan of the paper is as follows: section \ref{semi} describes and justifies the semiclassical approximation about a Minkowski background spacetime, section \ref{rel} discusses the properties of the relative entropy, section \ref{thermal} describes the thermal properties of the Rindler wedge, and section \ref{form} gives the proof of the GSL.  Finally, section \ref{dis} describes how to generalize the result to anti-de Sitter space and other spacetimes with Rindler-like horizons, and speculates how one might generalize the proof to arbitrary slices of arbitrary horizons.  I will use metric signature $(-,+,+,+)$ and $c = 1$, taking 4 dimensions for specificity.

\section{The Semiclassical Approximation}\label{semi}

Consider n-dimensional general relativity coupled to matter, described by the following action:
\begin{equation}\label{action}
I = \int d^4x (\sqrt{-g}\frac{R}{16\pi G} + \mathcal{L}_\mathrm{matter}).
\end{equation}
I will assume that the matter fields are minimally coupled to the metric, so that $\mathcal{L}_\mathrm{matter}$ does not lead to any additional corrections to the horizon entropy $S_H$.

\begin{figure}[ht]
\centering
\includegraphics[width=.9\textwidth]{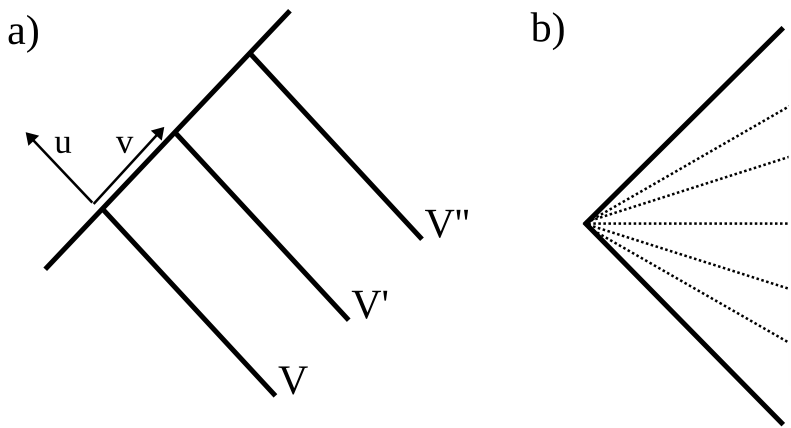}
\caption{\footnotesize a) The one parameter family of Rindler wedges in the $u$-$v$ coordinate system, illustrated by three particular wedges which share the same future Rindler horizon.  The wedges are related by null translations in the v direction.  The GSL states that each wedge should have at least as much generalized entropy as the wedges beneath it.  b) The boost symmetry of a single Rindler wedge, which is used to show that the vacuum state is thermal with respect to the boost energy.  The spatial slices related by the boost symmetry all have the same horizon area and the same entropy content, so the generalized entropy of each slice is constant, assuming there is no anomaly in the renormalization of the outside entropy.} \label{wedges}
\end{figure}

The equation of motion due to varying the metric is the Einstein equation
\begin{equation}
G_{ab} = 8\pi G\,T_{ab}
\end{equation}
where the matter stress-energy is defined as
\begin{equation}\label{SE}
T_{ab} = -\frac{2}{\sqrt{-g}}\frac{\delta \mathcal{L}_\mathrm{matter}}{\delta g^{ab}}.
\end{equation}
For $T_{ab} = 0$, one solution is the Minkowski vacuum, which can be written in null coordinates as follows:
\begin{equation}\label{null}
ds^2 = -2 du\,dv + dy^2 + dz^2.
\end{equation}
This spacetime has many Rindler horizons, but all of them are related by symmetry to the one defined by $u = 0$.  This Rindler horizon contains a 1-parameter family of Rindler wedges $W(V)$, defined as the locus of points satisfying
\begin{equation}\label{wedge}
u \le 0; \quad v \ge V,
\end{equation}
and the surface on which $u = 0$ and $v = V$ is called the bifurcation surface.  The wedge is invariant under a boost transformation whose Killing vector is given by
\begin{equation}
\xi = v \partial_v - u \partial_u.
\end{equation}
Note that if 
$V < V^\prime$, then $W(V) \supset W(V^\prime)$.  The GSL is now the statement that the generalized entropy $S_\mathrm{gen}(W(V)) \equiv S_\mathrm{gen}(V)$ should be a nondecreasing function of $V$.  Fig. \ref{wedges} shows how these wedges relate to one another.

In the semiclassical approximation around this Minkowski space background, $\mathcal{L}_\mathrm{matter}$ is regarded as the action for an ordinary quantum field theory (QFT).   This QFT should assign to each Rindler wedge $W(V)$ an algebra of observables $M(V)$, such that when $V < V^\prime$, $M(V) \supset M(V^\prime)$ (because every observable in the smaller wedge is also an observable of the larger one).  

The QFT should also have a renormalized stress-energy operator $T_{ab}$.  The semiclassical Einstein equation
\begin{equation}\label{semiEin}
G_{ab} = 8\pi G \langle T_{ab} \rangle
\end{equation}
determines the perturbation of the Minkowski space background (once boundary conditions are specified).  If the matter stress-energy is localized then the perturbed spacetime must remain asymptotically flat.  The Rindler wedge can still be defined on the perturbed spacetime as the the intersection of the future and the past of a uniformly accelerating worldline (or equivalently, the intersection of the future of a point on $\mathcal{I}^-$ with the past of a point on $\mathcal{I}^+$).  This definition can be made unambiguous even when the spacetime is gravitationally perturbed, by taking the accelerating observer to be very far from the matter, where spacetime is nearly flat.

For a state of the fields with characteristic wavelength $\lambda$ (in some inertial frame) and an order unity number of quanta, the expected stress-energy is of order $\hbar / \lambda^4$, which implies via the Einstein equation that the curvature is of order $\hbar G / \lambda^4$, and since the curvature involves two derivatives of the metric, the resulting metric perturbation is of order
\begin{equation}
a \equiv \hbar G / \lambda^2 = L_\mathrm{planck}^2 / \lambda^2.
\end{equation}
If $a$ is much less than unity it is possible to consistently neglect the effects of gravity on the quantum fields up to terms of order $a$.  Typically renormalization will induce nonminimal coupling terms into $\mathcal{L}_\mathrm{matter}$; however the effect of these terms on the canonically normalized metric are suppressed by positive powers of $a$.  Thus the only important effect of the metric perturbation is on the Bekenstein-Hawking area term, which affects $S_\mathrm{gen}$ by order one bit.  Similarly the changes in $S_\mathrm{out}$ should also be of order one bit.  Then the GSL states that when these two contributions are added together, the entropy $S(V)$ of the Rindler wedges $W(V)$ is a monotonic function of $V$.

The semiclassical approximation neglects the fluctuations in the metric.  These fluctuations appear for two reasons: first because of the quantization of gravitons, and second because the source term $T_{ab}$ has fluctuations.

\paragraph{\textbf{Graviton fluctuations.}}  Although gravitons carry canonical energy and momentum, they do not contribute to the matter stress-energy tensor $T_{ab}$ as defined in Eq. (\ref{SE}).  Nevertheless, $G_{ab}$ has terms which are quadratic in the metric, so in order to describe the equation of motion correctly when there are gravitons, it is necessary to quantize the metric field as well and impose $8\pi G T_{ab} = G_{ab}$ as an operator equation.  Schematically one can decompose the Einstein tensor in terms of the metric and derivatives as
\begin{equation}
\nabla^2 g + \nabla^2 g^2 + \mathcal{O}(\nabla^2 g^3),
\end{equation}
ignoring indices and what the derivatives act on.  One may now think of the metric as being decomposed into a) a background Minkowski metric, b) linearized gravity waves on top of this metric, and c) nonlinear effects, due to the fact that the Einstein tensor is nonlinear in the metric.  Although the linearized gravity waves do not contribute to $G_{ab}$ to first order, to second order they have a nonzero contribution due to the $\nabla^2 g^2$ terms; in fact the gravitons must contribute to the Einstein equation at the same order as ordinary matter quanta of the same wavelength.  In a state with an order unity number of gravitons, this contribution to the Einstein tensor goes like
\begin{equation}
\nabla^2 g^2 = \hbar G / \lambda^4 = a / \lambda^2,
\end{equation}
from which it follows that the amplitude of $g$ due to gravitons is of order $sqrt{a}$.  This second order contribution to the Einstein tensor is cancelled out by the nonlinear gravitational field which is induced by the linearized gravity waves, which is of order $a$.  Now in a state with a small number of quanta, the fluctuations in a field are of the same order as the field itself.  Thus graviton fluctuations are themselves of order $\sqrt{a}$---too large, in general, to be neglected.

Although I am confident that it is possible to generalize the proof below to the case in which there are gravitons, doing so would involve additional technical complications.  So in this paper I will restrict to states with zero gravitons in them.  Assuming that the past-boundary conditions include no gravitons, the amplitude for the matter fields to emit a graviton will be proportional to $\sqrt{a}$, as can be seen by canonically normalizing the metric field in Eq. (\ref{action}) and applying the usual Feynman rules.  Since the Einstein tensor depends quadratically on the graviton field, this means that the graviton contributions to the Einstein equation will be suppressed by an additional power of $a$ compared to the matter contributions, allowing them to be neglected.\footnote{Note that this argument depends on the fact that Minkowski space has a well-defined graviton vacuum state which evolves to itself under time evolution.  In contrast, if a black hole forms from collapse, there is in general Hawking radiation of gravitons, leading to an increase in the evaporation rate of the black hole which cannot be ignored.}

\paragraph{\textbf{Stress-Energy fluctuations.}}  For states with an order unity number of matter quanta, the quantum fluctuations in $T_{ab}$ are of the same order as the expectation value $\langle T_{ab} \rangle$, so it is not clear in general whether the semiclassical Einstein equation (\ref{semiEin}) is a good approximation.  These fluctuations in $T_{ab}$ cause fluctuations in the horizon entropy $A / 4\hbar G$ of order one bit.  However, given that the generalized entropy as defined in the Introduction depends only on the expectation value $\langle A \rangle$, these fluctuations do not affect the GSL as defined here, and can thus be ignored \cite{10proofs}.

\section{The Relative Entropy}\label{rel}

The relative entropy is an information-theoretic quantity which is closely related to the generalized entropy \cite{casini08}.  It satisfies a monotonicity property which will be used below to prove that the generalized entropy is increasing with time.  For any two density matrices $\rho$ and $\sigma$, the relative entropy is given by the formula
\begin{equation}\label{relI}
S(\rho\,|\,\sigma) = \mathrm{tr}(\rho\,\ln\,\rho) - \mathrm{tr}(\rho\,\ln\,\sigma).
\end{equation}
Intuitively speaking, the relative entropy measures how far away from each other two states $\rho$ and $\sigma$ are.  However, it is not a symmetric function of $\rho$ and $\sigma$.  In a system with $N$ different states, if $\sigma = 1/N$ (the uniformly mixed state), then the relative entropy is simply the difference between the entropies:
\begin{equation}
S(\rho\,|\,\sigma) = S(\sigma) - S(\rho) = \ln\,N + \mathrm{tr}(\rho\,\ln\,\rho).
\end{equation}
At the opposite extreme, when $\sigma$ is a pure state, then
\begin{eqnarray}
S(\rho\,|\,\sigma) = +\infty & \quad \mathrm{if} & \rho \ne \sigma, \\
S(\rho\,|\,\sigma) = 0 & \quad \mathrm{if} & \rho = \sigma.
\end{eqnarray}
In between these two cases, suppose that $\sigma$ is a Gibbs thermal equilibrium state with respect to some Hamiltonian,
\begin{equation}\label{Gibbs}
\sigma = \frac{e^{-\beta H}}{\mathrm{tr}(e^{-\beta H})}.
\end{equation}
Then Eq. (\ref{relI}) is equal to beta times the free energy difference of $\rho$ and $\sigma$:
\begin{equation}\label{Hafe}
S(\rho\,|\,\sigma) = 
[\beta\langle H \rangle_\rho -  S(\rho)] - [\beta\langle H \rangle_\sigma - S(\sigma)],
\end{equation}
using the definition of the von Neumann entropy $S(\rho) = -\mathrm{tr}(\rho\,\ln\,\rho)$, the fact that $\ln \sigma = -\beta H$ up to an additive constant, and the fact that the relative entropy vanishes when $\rho = \sigma$.

In any QFT, a regular state in a Rindler wedge has an infinite number of excited degrees of freedom residing near the horizon.  This implies that the definition of the relative entropy in Eq. (\ref{relI}) is ill-defined due to the inability to write the states $\rho$ and $\sigma$ as density matrices.  To see this, notice that the rows and columns of a density matrix ought to be labeled by a basis of pure quantum states.  But in the case of the Rindler wedge there are no pure states; the divergence in the entanglement entropy tells us that every physically acceptable state is mixed.\footnote{For readers familiar with algebraic QFT, the failure of Eq. (\ref{relI}) comes from the fact that the algebra of observables in any region with a boundary is actually a type III von Neumann algebra \cite{BAF87}, which by definition has no trace operation.}  A state $\rho$ can still be defined as a positive, normalized, linear functional 
$\rho(M)$ over some algebra of observables $M$.  Any such state defined on an algebra $M$ is automatically also a state of any subalgebra $M^\prime \in M$.

The relative entropy can still be defined for states in systems with an infinite number of degrees of freedom by taking a limit \cite{araki75}.  Let the system be described by a tensor product of an infinite number of Hilbert Spaces $\mathcal{H}_n$ where $n$ ranges over the natural numbers.  Then the relative entropy of the system is given by
\begin{equation}
\lim_{n \to \infty} (\mathrm{tr}(\rho_n\,\ln\,\rho_n) - \mathrm{tr}(\rho_n\,\ln\,\sigma_n)),
\end{equation}
where $\rho_n$ means $\rho$ viewed as a density matrix on the tensor product of the first $n$ Hilbert Spaces.  This is a special case of a more general definition which applies to arbitrary algebras of observables \cite{araki75}.

Some properties of the relative entropy: First of all, $S(\rho\,|\,\sigma)$ is always nonnegative, and is zero only when $\rho = \sigma$.  It may however take the value $+\infty$.  
More remarkably, the relative entropy is monotonic \cite{lindblad75}, meaning that whenever $\rho$ and $\sigma$ are restricted from one algebra (e.g. $M$) to a subalgebra (e.g. $M^{\prime}$), the relative entropy is nonincreasing:
\begin{equation}\label{mono}
S(\rho\,|\,\sigma)_M \ge S(\rho\,|\,\sigma)_{M^{\prime}}.
\end{equation}
Intuitively, when probed with fewer observables, $\rho$ and $\sigma$ are less distinguishable and therefore must have less relative entropy.  

This monotonicity property is reminiscent of the GSL.  My strategy for proving the GSL will be as follows: Let $\rho$ be the state which we wish to prove has nondecreasing entropy, and let $\sigma$ be the vacuum state, which is translation invariant with respect to the null coordinate $v$.  I will show that the generalized entropy is related to the relative entropy by
\begin{equation}\label{relgen}
S_\mathrm{gen}(\rho) = C - S(\rho\,|\,\sigma),
\end{equation}
where $C$ is a constant with respect to changes in the advanced-time null coordinate $v$.  Then the monotonicity of the relative entropy will imply the nondecrease of the generalized entropy.  So the entire burden of the proof that follows is to establish Eq. (\ref{relgen}) for each wedge $W(v)$.

The idea of relating the relative entropy to the generalized entropy is found in Casini \cite{casini08}, who shows how it is implicitly used in the quasi-steady proofs of the GSL due to Frolov \& Page \cite{FP93} and Sorkin \cite{sorkin98}.

\section{Thermal Properties of the Rindler Wedge}\label{thermal}

When the vacuum state $\sigma$ is restricted to a particular Rindler wedge $W(V)$ located at $v = V$, it is thermal with respect to the boost energy $K(V)$ conjugate to the boost symmetry of that wedge.  This is known as the Unruh effect, and has been proven for any QFT with a Lorentz symmetric ground state \cite{BW76}.  Technically this means that $\sigma$ satisfies the KMS condition \cite{KMS}: For any two observables $A$ and $B$, if $\alpha_{z}$ represents a Lorentz boost which translates observables by the hyperbolic angle $z$, $\langle B \alpha_{z}(A) \rangle_{\sigma}$ must be an analytic function of $z$ when $0 < \textrm{Im}(z) < i \hbar \beta$, and also
\begin{equation}
\langle AB \rangle_{\sigma} = \langle B \alpha_{i \hbar \beta}(A) \rangle_{\sigma},
\end{equation}
where $\beta = 2\pi / \hbar$ is the inverse Unruh temperature.

The boost energy associated with the wedge $W(V)$ is defined as the following integral of the stress-energy tensor over any complete time slice $\Sigma$ stretching from the bifurcation surface to infinity:
\begin{equation}\label{boost}
K = \int_\Sigma T_{ab} \xi^a d\Sigma^b
\end{equation}
where $\xi^a$ is the Killing vector of the boost symmetry, and 
\begin{equation}
d\Sigma^a = \sqrt{-g} g^{ae} \epsilon_{ebcd}
\end{equation}
is a vector-valued 3-form obtained from the metric and the permutation symbol.  In principle, one should find $K$ by integrating the canonical stress-energy tensor derived from Noether's theorem, rather than the gravitational stress energy tensor $T_{ab}$ found by varying the metric.  That is because the canonical boost energy is the generator of the boost symmetry of the Rindler wedge.  However, in the case of minimally coupled fields the canonical and gravitational stress-energies are the same (e.g. \cite{fursaev99}), so the use of the gravitational stress-energy tensor in Eq. (\ref{boost}) is correct.

Since the KMS state is thermal in the boost energy, Eq. (\ref{Hafe}) suggests that the relative entropy of a state $\rho$ to the vacuum state $\sigma$ can be written as a difference of free boost energies:
\begin{equation}\label{afe}
S(\rho\,|\,\sigma) = \beta \langle K \rangle_\rho - S_\mathrm{out}(\rho) + S_\mathrm{out}(\sigma),
\end{equation}
where a $\langle K \rangle_{\sigma}$ term need not be included because the renormalized stress-energy vanishes in the vacuum.  However, this formula was only derived above for systems  described by a Hilbert Space, and does not apply to Rindler space.  Because of this, $\sigma$ is only formally a Gibbs state $e^{-\beta K}/\mathrm{tr}(e^{-\beta K})$, and some other justification is needed to rigorously show Eq. (\ref{afe}).

Since the outside entropy $S_\mathrm{out}$ is divergent and therefore needs some procedure for making it well-defined, one could take Eq. (\ref{afe}) as the \emph{definition} of $S_\mathrm{out}(\rho)$ (up to a constant) for all states with a finite value of $K$.  But this is unsatisfying because it is unclear that $S_\mathrm{out}(\rho)$ defined in this manner satisfies the expected properties of the entropy, such as being invariant under unitary transformations of $\rho$ inside the wedge, or being equal on both sides of the horizon when the total state is pure.

It is very plausible that Eq. (\ref{afe}) holds for the Rindler Wedge in QFT.  For example, an analogue of this result has been shown for infinite quantum spin-systems by Araki and Sewell (Eq. (2.15) in Ref. \cite{AS77}).  The conventional wisdom is that any QFT can be discretized on a lattice, which strongly suggests that a corresponding statement should also hold for an arbitrary QFT.  However the justification of Eq. (\ref{afe}) is tied up in the difficult question of how to rigorously renormalize the generalized entropy.  Here I will simply assume that Eq. (\ref{afe}) holds when when $S_\mathrm{out}$ is interpreted as the renormalized entanglement entropy.

\section{The Generalized Entropy Increases}\label{form}

In this section it will be shown that the generalized entropy $S(v)$ associated with the wedges $W(v)$ is a nondecreasing function of $v$, by relating it to the relative entropy to the vacuum state $\sigma$.

\begin{figure}[ht]
\centering
\includegraphics[width=.75\textwidth]{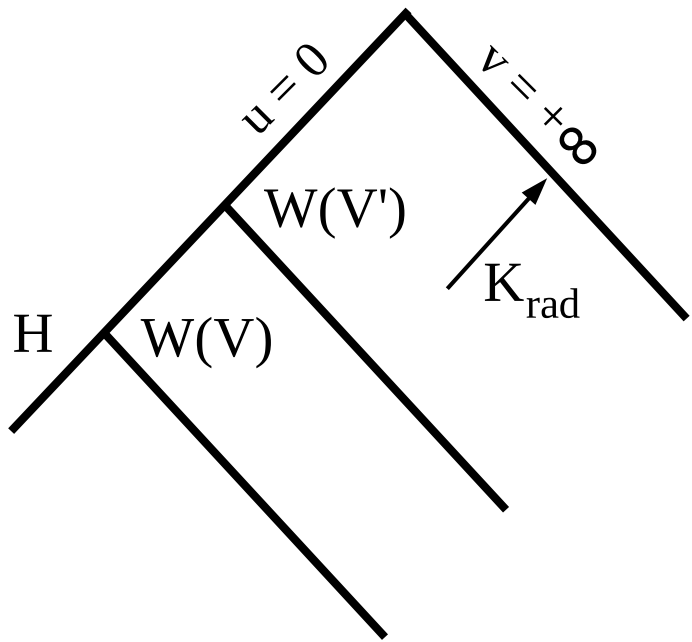}
\caption{\footnotesize The wedge $W(V)$ evolves forward in time to $W(V^\prime)$.  Each of the wedges contains a certain amount of boost energy $K$ all of which must either fall across the horizon $H$ or be radiated to infinity and thus contribute to $K_\mathrm{rad}$.  The total amount of boost energy in each wedge is thus proportional to the area of the wedge, up to the contribution at $v = +\infty$, which is the same for both $W(V)$ and $W(V^\prime)$.} \label{increase}
\end{figure}

Consider one particular wedge $W(V)$ at time $v = V$ on the horizon defined by $u = 0$.  The boost energy $K(V)$ is given by Eq. (\ref{boost}) for all complete time slices.  Choose the slice $\Sigma$ to be the future horizon $H$ itself plus the asymptotic null future $v = +\infty$ as shown in Fig. \ref{increase}.  The boost energy is now given by the following integral on $H$:
\begin{equation}\label{K}
K(\rho) = \int_{H;\,v > V} T^{uu}(v - V) dv\,d^2x + K_\mathrm{rad},
\end{equation}
where $d^2x$ represents the integration over the two spacelike horizon directions, and $K_\mathrm{rad}$ is the total amount of boost energy which radiates to null infinity instead of falling across the horizon.\footnote{In a generic state, $K_\mathrm{rad}$ equals zero, because the only way for a particle not to fall across the Rindler horizon is to travel away at the speed of light in the direction exactly perpendicular to the horizon.  But this consideration does not apply to black hole horizons, from which generic matter can escape to infinity.}  This radiated boost energy is given by
\begin{equation}\label{Krad}
K_\mathrm{rad} = \int_{v = +\infty;\,u < 0} T^{vv} (-u) du\,d^2x.
\end{equation}
By virtue of conservation of boost energy, the $v \to +\infty$ limit needed to define Eq. (\ref{Krad}) is well-defined in any state that has a finite amount of boost energy falling across the horizon and coming in from past null infinity.  Since $K_\mathrm{rad}$ is not a function of $v$, it is the same for each wedge $W(V)$ and therefore does not contribute to the change in the generalized entropy with time.

When gravitational interactions are taken into account, the boost energy falling across the horizon leads to a small, order $a$ semiclassical correction in the area of the bifurcation surface of the wedge $W(V)$.  The linearized Raychaudhuri equation, together with the Einstein equation, says that
\begin{equation}\label{linray}
\frac{d \theta }{dv} = -8\pi G\,T_{ab} k^a k^b.
\end{equation}
where $k^a = g^{ab} u_{,b}$, and $\theta = (1/A)(dA/dv)$ is the expansion.\footnote{Strictly speaking, Eq. (\ref{linray}) is only justified for the region of the horizon which is not too far to the past of the quantum matter perturbation.  That is because the matter fields will cause the horizon generators to focus, meaning that going backwards in time, the horizon generators will eventually form cusps and leave the event horizon altogether.  Near these cusps, the geometry of the horizon cannot be treated as a small perturbation, since even though the metric fluctuations are small, the horizon location has large fluctuations.  However, the nonlinearities in the Raychaudhuri equation only make the horizon area increase faster with time, so the GSL should also hold in this region.  See Refs. \cite{JP07}\cite{AMV08} for the related issue of applying the first law to Rindler horizons.}  The $\theta^2$ and $\sigma_{ab}\sigma^{ab}$ terms are of order $a^2$ and can therefore be neglected.\footnote{However, in situations where one must take into account gravitons, there are $\sqrt{a}$ metric perturbations as described in section \ref{semi}.  This would make the $\sigma_{ab}\sigma^{ab}$ also of order $a$.  To adapt the proof to this circumstance, one would have to include the contribution of the gravitons themselves to the boost energy $K$.}  At advanced time $v = +\infty$, the future horizon should be unaffected by any stress-energy, and should therefore obey the stationary boundary condition
\begin{equation}
\theta|_{v=+\infty} = 0.
\end{equation}
Using this boundary condition, one may solve for the area of the bifurcation surface of $W(V)$ by integrating Eq. (\ref{linray}) twice along the $v$ direction and once along each spacelike dimension of the future horizon:
\begin{eqnarray}
A(V)= A(\infty) - 8\pi G \int_{H;\,v > V} T_{ab} \, k^a k^b (v - V) dv\,d^2x \\
= A(\infty) - 8\pi G K(V) - K_\mathrm{rad},
\end{eqnarray}
where I am suppressing expectation value signs.  (It makes no difference whether one performs these integrals on the perturbed or unperturbed horizons.  Because the integrand is already of order $a$, the error from integrating on the unperturbed horizon is of order $a^2$.)  This establishes Eq. (\ref{AK}), showing that the horizon area is equal to the boost energy up to an additive constant.  Note that because $v - V = 0$ on the bifurcation surface, the instantaneous boost energy change $dK/dV$ is entirely due to changes in the boost Killing vector $\xi$ used to define $K$, rather than due to any boost energy falling across the horizon at the bifurcation surface.

One can now apply Eq. (\ref{afe}) in order to write $A(V)$ in terms of the relative entropy,
\begin{equation}
\langle A(V) \rangle = \langle A(\infty) \rangle + 8\pi G \langle K_\mathrm{rad} \rangle - \frac{8\pi G}{\beta}[S(\rho\,|\,\sigma) + S_\mathrm{out}(\rho) - S_\mathrm{out}(\sigma)]_V
\end{equation}
But the final horizon area $A(\infty)$, the null energy radiated to infinity $K_\mathrm{rad}$, and the renormalized entanglement entropy of the vacuum $S_\mathrm{out}(\sigma)$ are all constants with respect to the advanced time $V$.  Setting $\beta = 2\pi / \hbar$, one finds that
\begin{equation}\label{final}
-S(\rho\,|\,\sigma) = 
S_\mathrm{out} + \langle A \rangle /{4\hbar G} = S_\mathrm{gen}(\rho) + \mathrm{const.},
\end{equation}
Then the monotonicity of the relative entropy implies that the generalized entropy is nondecreasing.

\section{Discussion}\label{dis}

The above result shows that any QFT minimally coupled to Einstein gravity obeys the GSL semiclassically for Rindler horizons.  The proof assumes that some suitable renormalization scheme exists which validates the formal relation (\ref{afe}) between the relative entropy, the outside entropy, and the boost energy.  This extends the proof of the GSL to rapidly changing quantum fields.

To summarize the proof: the area is related to the boost energy by means of Eq. (AK):
\begin{equation}
A = \mathrm{const.} - 8\pi G K.
\end{equation}
This is related to the fact that in general relativity the horizon area is canonically conjugate to the Killing time \cite{CT93}.  The generalized entropy can then be written out in terms of the free boost energy with $\beta = 2\pi / \hbar$:
\begin{equation}
S_\mathrm{gen} = \mathrm{const.} - \beta K + S_\mathrm{out}.
\end{equation}
But the free boost energy is related to the relative entropy
\begin{equation}
\beta K - S_\mathrm{out} = \mathrm{const.} + S(\rho\,|\,\sigma),
\end{equation}
and since the relative entropy can never increase, the generalized entropy can never decrease.

I have assumed above that the background spacetime is Minkowski.  This restriction can actually be lifted somewhat, to any spacetime with an infinite 1-parameter family of nested wedges 
$W(v)$, such that each wedge has a positive boost Killing field.  Since the commutator of any two boosts is a null translation on the horizon, these symmetries generate a 2-dimensional Lie group of null translations and boosts of the future horizon.  Choosing coordinates $(u, v, x^i)$ on the spacetime with the property that this group acts in the standard way,
\begin{eqnarray}
v &\to& av + b, \\
u &\to& u/a,
\end{eqnarray}
the most general possible resulting spacetime is the following metric:
\begin{eqnarray}\label{metric}
& ds^2 = - f(x^i)\,du\,dv - g(x^i) u^2\,dv^2 + 
h_a(x^i) u\,dv\,dx^a + q_{ab} (x^i) \,dx^a \,dx^b, &\\
& f > 0, \quad q_{ab} = \mathrm{pos.\,def.}, \quad g \ge 0, &
\end{eqnarray}
where the first two constraints are necessary to ensure a Lorentzian signature, and the third is necessary for the boost Killing vector to be future timelike inside each wedge $W(v)$.  The condition $g \ge 0$ automatically also implies that the translation Killing vector is future-null or future-timelike everywhere.  Hence in a stable theory there should exist a ground state $\sigma$ of the null- translation symmetry.  This implies that $\sigma$ is a KMS state with respect to each of the boost Killing vectors \cite{sewell81}, and is translation-invariant.  This is all that is needed for the argument in section \ref{form}, so the GSL must hold on these spacetimes too.

Metrics of the form Eq. (\ref{metric}) include anti-de Sitter space or Minkowski space tensored with any other Riemannian geometry.\footnote{Of course, if the spacetimes are not Ricci-flat it is necessary to postulate classical background matter fields sourcing the Ricci tensor.  The proof would then apply to quantum perturbations of such spacetimes.}  However, neither de Sitter space nor black hole spacetimes qualify, because neither spacetime has a Killing vector which points to the future everywhere.  This means than except on the bifurcation surface, there is no analogue of the boost-symmetric thermal Rindler wedge.  Since my proof requires both the initial and final outside regions to be thermal, it does not apply to such spacetimes.

Even in qualifying spacetimes, the result here only shows the GSL for those slices of the horizon which are bifurcation surfaces.  Otherwise there is no boost symmetry of the exterior region outside of the slice, and hence no thermal state.  But on a fully dynamical horizon there are no approximate bifurcation surfaces, so if the GSL applies to such horizons there would have to exist a more local version of the GSL which would apply to arbitrary slices of the horizon.  This more local version of the GSL would imply other important results such as the averaged null energy condition \cite{anec}.

Both the horizon restrictions and the slice restrictions might be overcome by invoking some sort of near-horizon limit, by exploiting the fact that for an arbitrary horizon slice, there is an \emph{approximate} boost symmetry very close to the horizon slice, which guarantees that the fields are approximately thermal very close to the horizon.  Furthermore, there is an approximate null translation symmetry relating any two nearby slices locally.  Assuming that the question of whether or not entropy increases comes down to what happens very close to the horizon, the GSL could then be shown for arbitrary horizons.  The challenge of such an approach would be to find a helpful way to take advantage of the near-horizon limit despite the fact that thermodynamic quantities like $S_\mathrm{out}$ are defined globally on the entire exterior region.  Such an approach might follow Ref. \cite{sewell81}, in which the thermality of a Schwarzschild black hole is a consequence of a null translation symmetry of the horizon, despite the fact that this symmetry does not extend to the rest of the spacetime.

Another limitation of the present result is the restriction to fields which are minimally coupled to general relativity.  This assumption came into the proof in two different ways:  
1) in the assumption that the horizon entropy is $A/{4\hbar G}$, rather than the Wald entropy defined by differentiating the Lagrangian with respect to the Riemann tensor \cite{WI94}, and 2) in the assumption that the Rindler wedge is thermal with respect to the boost energy derived from the gravitational stress-energy tensor $T_{ab}$, rather than the canonical boost energy.  Classically, the difference between the canonical and gravitational stress-energies is simply proportional to the contribution of the matter fields to the Wald entropy \cite{fursaev99}, so these two errors probably cancel out, so that the GSL still holds.  Since the canonical boost energy includes contributions from gravity waves, such a proof might also automatically apply to states containing gravitons.  But in order to show this rigorously, it would be necessary to show that these properties of the Wald entropy hold even when metric perturbations are quantized.

\small
\subsection*{Acknowledgments}
I am grateful for comments from Ted Jacobson and William Donnelly.  Supported in part by the National Science Foundation grants PHY-0601800 and PHY-0903572, and the Maryland Center for Fundamental Physics.

\end{document}